\documentclass[reprint,aps,ams,amsmath,prx,twocolumn,superscriptaddress,longbibliography]{revtex4-1}
\usepackage{graphics}
\usepackage{graphicx}
\usepackage{epsfig}
\usepackage{amsmath}
\usepackage{amssymb}
\usepackage{amsfonts}
\usepackage{bm}
\usepackage{mathrsfs}
\usepackage{color}
\usepackage{bbm}
\usepackage{hyperref}
\usepackage[normalem]{ulem}
\usepackage{comment}
\usepackage{soul}

\newcommand{\be}{\begin{equation}}
\newcommand{\ee}{\end{equation}}
\def\rr#1{(\ref{#1})}

\begin{document}

\title{Fluctuation-induced giant  magnetoresistance in charge-neutral graphene}

\author{A. Levchenko}
\affiliation{Department of Physics, University of Wisconsin-Madison, Madison, Wisconsin 53706, USA}

\author{E. Kirkinis}
\affiliation{Center for Computation and Theory of Soft Materials, Robert R. McCormick School of Engineering and Applied Science, Northwestern University, Evanston IL 60208 USA}

\author{A. V. Andreev}
\affiliation{Department of Physics, University of Washington, Seattle, Washington 98195, USA}

\date{\today}

\begin{abstract}
The Johnson-Nyquist noise associated with the intrinsic conductivity of 
the electron liquid, induces fluctuations of the electron density in charge-neutral graphene devices. In the presence of external electric and magnetic fields, the fluctuations of charge density and electric current induce a fluctuating hydrodynamic flow. We show that the resulting advection of charge produces a fluctuation contribution to the macroscopic conductivity of the system, $\sigma_{\mathrm{fl}}$, and develop a quantitative theory of $\sigma_{\mathrm{fl}}$. At zero magnetic field,  $\sigma_{\mathrm{fl}}$ diverges logarithmically with the system size and  becomes rapidly suppressed at relatively small fields. This results in giant magnetoresistance of the system. 
\end{abstract}

\maketitle


It has been appreciated 
since the pioneering work of Gurzhi~\cite{Gurzhi:1968} that charge transport in high-mobility conductors at finite temperature involves the hydrodynamic flow of the electron liquid. 
The hydrodynamic regime is realized in the range of temperatures and sample purities, in which 
the momentum relaxation length caused by electron-impurity and electron-phonon scattering exceeds the relaxation length due to momentum conserving electron-electron collisions. In this case, the resistance of the system is determined by the viscosity and thermal conductivity of the electron liquid (see, e.g., the reviews \cite{Spivak:2010,NGMS:2017,Lucas:2018,Schmalian:2020,Narozhny:2022,Scaffidi:2024}, as well as Refs. \cite{Andreev:2011,Levchenko:2017}).

An important exception to the coupling between charge transport and hydrodynamic flow occurs in compensated conductors, such as graphene at charge neutrality. In this case, the hydrodynamic flow of the charge-neutral electron liquid corresponds to the flow of heat, while the electric current is mediated by the intrinsic conductivity of the liquid \cite{Mishchenko:2007,Kashuba:2008,Fritz:2008}. 
Because of this, the macroscopic conductivity extracted from transport measurements at charge neutrality is typically interpreted as the intrinsic conductivity of the electron liquid \cite{Kim:2007,Du:2008,Chen:2008,Stormer:2008,Mayorov:2012,Crossno:2016,Morpurgo:2017,Gallagher:2019}. 

In this work, we show that this interpretation requires revision. The decoupling between the electric current and the hydrodynamic flow holds only on average. Thermal fluctuations inevitably generate local deviations from charge neutrality and induce a coupling between the electric current and the hydrodynamic velocity field. In what follows, we develop a theory of hydrodynamic fluctuations in charge-neutral electron liquids taking into account this coupling.  

It is well known~\cite{Ernst:1971,Forster:1977,Andreev:1980}, that in two-dimensional (2D) liquids the  hydrodynamic fluctuations produce a contribution to the shear viscosity, which diverges logarithmically with system size. We find that for liquids with a nonvanishing intrinsic conductivity $\sigma_0$, the contribution of the hydrodynamic fluctuations to the conductivity also diverges logarithmically with the size of the system. This  contribution is very sensitive to magnetic field and dominates the magnetoresistance of charge-neutral graphene across a broad range of magnetic fields.

\begin{figure}[t!]
\includegraphics[width=\linewidth]{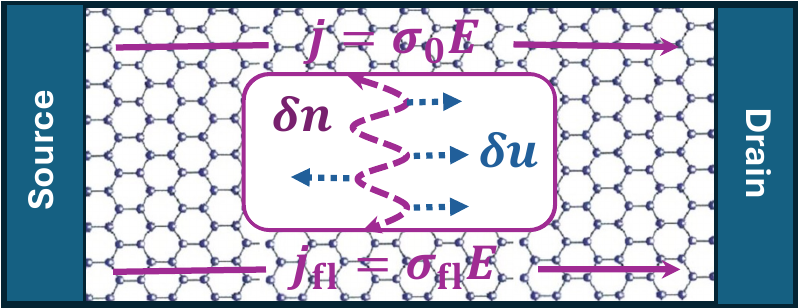}
\caption{Illustration of the fluctuation conductivity mechanism for a graphene Hall-bar device.  The Johnson–Nyquist noise, Eq.~\eqref{eq:Langevin}, 
generates spatial and temporal carrier-density fluctuations $\delta n_{\bm{q},\omega}$, shown by wavy lines. Under an applied electric field $\bm{E}$, this induces inhomogeneous hydrodynamic velocity $\delta\bm{u}_{\bm{q},\omega}$, which is correlated with $\delta n_{\bm{q},\omega}$. The resulting advection of charge produces  enhancement of the macroscopic conductivity, Eq.~\eqref{eq:DCj}.} 
\label{fig:Hall-bar}
\end{figure}

The physical mechanism underlying the fluctuation-driven enhancement of the intrinsic conductivity can be  understood as follows. Consider a graphene Hall-bar device shown in Fig. \ref{fig:Hall-bar}. The Johnson–Nyquist noise produces  fluctuations of the electron density
$\delta n$. The force exerted by an applied electric field $\bm{E}$ on the fluctuating charge density induces a  modulation of the hydrodynamic velocity  
$ \delta \bm{u}$ whose magnitude is determined by the balance between viscous force and the driving force. For the Fourier amplitude with a wavevector $q$, this yields $\eta q^2  \delta \bm{u}_q \sim e\bm{E} \delta n_q$,  where $\eta$ is the  shear  viscosity. The correlation between the velocity and charge density fluctuations, creates an average advection current proportional to $\bm{E}$, which enhances the conductivity.
For Coulomb interaction screened by a gate, charge density fluctuations decay at a rate $\gamma_q\propto q^2$. The average advection current caused by the fluctuations can then be estimated as $\bm{j}_{\text{fl}}\simeq e^2\bm{E}\sum_q\langle\delta n^2_q\rangle/(\eta q^2\gamma_q)$. The thermal average of the density fluctuations follows from the fluctuation–dissipation theorem, $\langle\delta n^2_q\rangle\propto T\sigma_0q^2$, where $\sigma_0$	
is the intrinsic conductivity of the pristine electron liquid. It follows from these estimates that the  fluctuation-induced  conductivity diverges logarithmically at long distances. Below we develop a quantitative theory of the fluctuation contribution, $\sigma_{\mathrm{fl}}$, to the system conductivity and its dependence on the magnetic field. We find that the magnetoresistance of the system is dominated by the fluctuation conductivity in a wide interval of the magnetic fields.   

The influence of hydrodynamic fluctuations on charge transport can be studied by applying the method developed by Landau and Lifshitz for Newtonian liquids \cite{LL:1957,LL-V9,Forster} to the case of charge-neutral electron liquids with a nonvanihshing intrinsic conductivity. To this end, Langevin sources are introduced into the hydrodynamic equations~\cite{Levchenko:2022}. We find that the dominant fluctuation corrections to the electrical conductivity arise from the Johnson-Nyquist noise associated with the intrinsic conductivity of the electron liquid. Therefore, below we focus on this contribution and defer the comprehensive treatment of hydrodynamic fluctuations in charge-neutral electron liquids to a separate work.

The starting point of our treatment is the continuity equation for the charge current
\begin{equation}\label{eq:continuity}
e\partial_t n+\bm{\nabla}\cdot\bm{j}=0
\end{equation}
that relates fluctuations of electron particle density $n(\bm{r},t)$ to current density $\bm{j}(\bm{r},t)$, whereby
\begin{equation}\label{eq:j}
\bm{j}=\sigma_0\bm{\mathcal{E}}+en\bm{u}+\delta\bm{j}, 
\end{equation}
$\bm{u}$ is the hydrodynamic velocity and $\sigma_0$ is the intrinsic conductivity, which is finite in systems lacking Galilean invariance. The last term in Eq. \eqref{eq:j} is the random Langevin current whose correlation function is given by the fluctuation-dissipation relation 
\begin{equation}\label{eq:Langevin}
\langle \delta \bm{j}_\alpha(\bm{r},t)\delta \bm{j}_\beta(\bm{r}',t') \rangle=2T\sigma_0\delta_{\alpha\beta}\delta(\bm{r}-\bm{r}')\delta(t-t'). 
\end{equation} 
For the electron liquid subjected to the crossed electric and magnetic fields, the electromotive force (EMF) is given by 
\begin{equation} \label{eE}
e\bm{\mathcal{E}}=e\bm{E}-\bm{\nabla}(\mu+e\phi)+\frac{e}{c}[\bm{u}\times\bm{H}]. 
\end{equation}
Here $c$ is the speed of light, $\phi$ is the electric potential related to the
electron density by the Poisson equation, and $\mu$ is the local chemical potential (we work in units $k_B=\hbar=1$). The last term in \rr{eE}
describes the contribution of the Lorentz force to the EMF, exerted by a magnetic field $\bm{H}$ on
the moving liquid \cite{LL-V8}. The linearized momentum evolution equation is given by
\begin{align}\label{eq:p}
&\rho\partial_t\bm{u}=en\bm{E}-s\bm{\nabla}T+\frac{1}{c}[\bm{j}\times\bm{H}]+\eta\bm{\nabla}^2\bm{u}+\bm{\nabla}\cdot\hat{\bm{\xi}}, 
\end{align}    
where $s$ is the entropy density, $\rho$ is the mass density, and $\eta$ is the shear viscosity of electron liquid. The last term describes the momentum flux due to the random Langevin stress tensor $\hat{\bm{\xi}}\equiv \xi_{ij}$. In what follows, we neglect the forces generated by the fluctuating viscous stresses since they contain one additional power of spatial gradients in comparison to the Lorentz force of the Langevin current and therefore become subleading at long wavelengths.
 
We use the Fourier representation and denote the fluctuating fields by their respective Fourier components: the particle density $n_{\bm{q},\omega}$ and the hydrodynamic velocity $\bm{u}_{\bm{q},\omega}$.
Inserting $\bm{j}$ from \eqref{eq:j} into \eqref{eq:continuity} one finds at charge neutrality
\begin{align}\label{eq:dndt}
\sigma_0\left(\frac{q^2}{e}(U(q)+\partial_n\mu)n_{\bm{q},\omega}+\frac{i}{c}\bm{H}\cdot[\bm{q}\times\bm{u}_{\bm{q},\omega}]\right)\nonumber \\ 
=i\omega en_{\bm{q},\omega}-i\bm{q}\cdot\delta\bm{j}_{\bm{q},\omega}, 
\end{align}
where the fluctuating part of the EMF was related to the Coulomb potential $U(q)$ and the compressibility of the fluid $\partial_n \mu$, which is the inverse of the thermodynamic density of states. 

At small velocities the liquid is incompressible, and  
the main enhancement of the conductivity  arises from the transverse velocity fluctuations~\footnote{This is similar to the enhancement of  conductivity of charge-neutral graphene \cite{Li:2020} in the presence of puddle disorder, which is induced by vortical flow.  }.
This enables us to neglect the  fluctuations of temperature because they affect only the longitudinal components of the velocity in Eq. \eqref{eq:p}.
From  Eq.~\eqref{eq:p} for the transverse components of the velocity fluctuations one obtains 
\begin{align}\label{eq:dpdt}
e n_{\bm{q},\omega}\left(\bm{E}-\frac{\bm{q}(\bm{q}\cdot\bm{E})}{q^2}\right)-\frac{\sigma_0H^2}{c^2}\bm{u}_{\bm{q},\omega}-\eta q^2\bm{u}_{\bm{q},\omega}\nonumber \\ 
=-i\omega\rho\bm{u}_{\bm{q},\omega}-\frac{1}{c}[\delta\bm{j}_{\bm{q},\omega}\times\bm{H}].
\end{align}

The solution of the linear system of Eqs.~\eqref{eq:dndt} and \eqref{eq:dpdt} is
\begin{subequations}
\label{eq:fluct_solution}
\begin{align}
&en_{\bm{q},\omega}=\frac{-i(\bm{q}\cdot\delta\bm{j}_{\bm{q},\omega})(\gamma_\nu-i\omega)}{(\gamma_\sigma-i\omega)(\gamma_\nu+\gamma_H-i\omega)+\frac{i\sigma_0}{c\rho}\bm{q}\cdot[\bm{E}\times\bm{H}]}, \\ 
&\bm{u}_{\bm{q},\omega}=\frac{[\hat{\bm{z}}\times\bm{q}](\bm{q}\cdot\delta\bm{j}_{\bm{q},\omega})(i\bm{q}[\hat{\bm{z}}\times\bm{E}]-\frac{H}{c}(\gamma_\sigma-i\omega))}{\rho q^2((\gamma_\sigma-i\omega)(\gamma_\nu+\gamma_H-i\omega)+\frac{i\sigma_0}{c\rho}\bm{q}\cdot[\bm{E}\times\bm{H}])}. 
\end{align}
\end{subequations}
where 
\begin{equation}
\gamma_\sigma=\frac{\sigma_0}{e^2}(U(q)+\partial_n\mu)q^2,\quad \gamma_\nu=\nu q^2,\quad \gamma_H=\sigma_0H^2/c^2\rho. 
\end{equation}
Here we introduced the following characteristic energies in the problem:  $\gamma_\sigma$ describes attenuation of fluctuations due to the Maxwell mechanism of charge relaxation, similarly $\gamma_\nu$, with $\nu=\eta/\rho$ being kinematic viscosity, describes relaxation via viscous decay of vorticity. Finally, $\gamma_H$ describes attenuation due to magnetic friction \cite{Muller:2008a}.
The fluctuation-induced contribution to the DC current density  is then given by 
\begin{equation}\label{eq:DCj}
\bm{j}_{\text{fl}}=\int\frac{d^2qd\omega}{(2\pi)^3}\langle en_{-\bm{q},-\omega}\bm{u}_{\bm{q},\omega}\rangle=\sigma_{\text{fl}} \bm{E},
\end{equation}
where the average $\langle\ldots\rangle$ is taken over the Langevin fluxes with the correlation function from Eq. \eqref{eq:Langevin}. It defines the fluctuation-induced contribution, $\sigma_\mathrm{fl}$,  to the macroscopic conductivity.  The total conductivity is thus given by $\sigma(T,H)=\sigma_0+\sigma_{\text{fl}}$.

\begin{figure}[t!]
 \includegraphics[width=\linewidth]{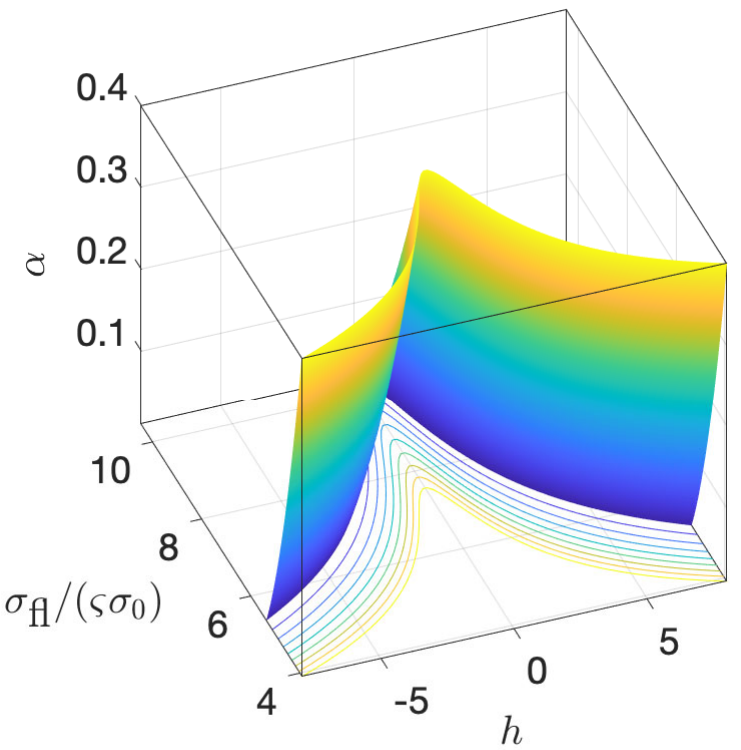}
 \caption{Magnetic field dependence of the fluctuation-induced conductance of charge neutral graphene $\sigma_{\text{fl}}$ corresponding to the main result of Eq. \eqref{eq:sigma-H} plotted for different values of the parameter $\alpha$ (denoted by the varicolored pattern) defined in Eq. \eqref{eq:H_T}.}
 \label{fig:sigma-H}
 \end{figure}

Substituting Eq.~\eqref{eq:fluct_solution} into Eq.~\eqref{eq:DCj}, and performing the average with the help of \eqref{eq:Langevin},  evaluating the frequency integral  and retaining the terms that are linear in $\bm{E}$, we obtain the fluctuation conductivity in the form 
\begin{equation}
\sigma_{\text{fl}}=\int\frac{d^2q}{(2\pi)^2}\frac{T\sigma_0q^2}{4\rho}\frac{\gamma_\sigma(\gamma_\nu+\gamma_H)+\gamma^2_\nu}{\gamma_\sigma(\gamma_\nu+\gamma_H)(\gamma_\sigma+\gamma_\nu+\gamma_H)^2}.
\end{equation}      
In the presence of a gate located at a distance $d$ from the sample, which was shown recently to enhance the electronic quality of graphene devices \cite{Domaretskiy:2025}, the interaction potential is screened 
\begin{equation}
U(q)=\frac{2\pi e^2}{q}\left(1-e^{-2qd}\right). 
\end{equation} 
Working under the assumption that $d$ is shorter than the thermal de Broglie length, $d\lesssim\lambda_T$, one can approximate $U(q)\approx 4\pi e^2d$. As a result of final momentum integration, one finds 
\begin{align}\label{eq:sigma-H}
\frac{\sigma_{\text{fl}}}{\sigma_0}= &\,\varsigma\left[\frac{2}{\alpha+1}\ln\frac{L}{l_{ee}}+\frac{\alpha(\alpha+1)-1}{\alpha+1}\ln\left(1+\frac{h^2}{1+\alpha}\right)\right.\nonumber \\ 
&\left.-\alpha\ln\left(1+\frac{h^2}{\alpha}\right)+\frac{(\alpha-1)h^2}{(\alpha+1)(\alpha+1+h^2)}\right].
\end{align}
Here, the logarithmic divergence of the momentum integration was cut off by the finite electron mean free path in the ultraviolet, $q_{\mathrm{max}}\sim 1/l_{ee}$, and by the system size in the infrared $q_{\mathrm{min}}\sim1/L$. The dimensionless factor $\varsigma$ is given by 
\begin{equation}
\label{eq:varsigma}
\varsigma=\left(\frac{e^2}{\sigma_0}\right)^2\frac{T \left(\partial_\mu n \right)^2}{4\rho}\, f^2(2\varkappa d), \quad f(x)=\frac{1}{1+x},
\end{equation} 
where we introduced the inverse Thomas-Fermi screening radius $\varkappa=2\pi e^2\partial_\mu n$. We also introduced the dimensionless magnetic field $h$ and the dimensionless parameter of the model $\alpha$ through the following relations 
\begin{equation} \label{eq:H_T}
h=\sqrt{\frac{(\partial_\mu n) L^2 f(2\varkappa d)}{4\pi^2\rho l^4_H}}=\frac{H}{H_T},\quad 
\alpha=\frac{e^2\nu(\partial_\mu n)}{\sigma_0}f(2\varkappa d),
\end{equation}
where $l_H=\sqrt{c/eH}$ is the magnetic length. The functional form of Eq. \eqref{eq:sigma-H} is plotted in Fig. \ref{fig:sigma-H} for a range of parameters in the model. It should be noted that the hydrodynamic equations used above are applicable only when the magnetic field $H$ is not too strong, such that Landau quantization of thermal excitations can be neglected. This requirement may be expressed as a condition on the cyclotron frequency $\omega_c$. The same condition, $\omega_c\tau_{ee}\ll1$, ensures that the magnetic-field dependence of both the intrinsic conductivity and the kinematic viscosity can also be ignored.

Let us consider the temperature-dependence of the parameters  $\varsigma$, $\alpha$ and $h$ in Eqs.~\eqref{eq:varsigma} and \eqref{eq:H_T} for monolayer (MLG) and bilayer (BLG) graphene devices. These parameters control the width and shape of the magnetoconductivity  in Eq.~\eqref{eq:sigma-H}. 

For MLG, the dispersion is linear $\epsilon=vp$. For BLG it is quadratic $\epsilon=p^2/2m$, and the effective mass can be estimated as $m\sim \Delta/v^2$, where $\Delta$ is the gap to the higher subbands. 
The thermodynamic density of states, 
\[
\partial_\mu n=-4\int\frac{d^2p}{(2\pi)^2}\frac{dn_F(\epsilon)}{d\epsilon}, 
\]
with  $n_F(\epsilon)=1/(e^{\epsilon/T}+1)$ being the Fermi function, is given by $\partial_\mu n=T\ln 4/(\pi v^2) $ for the monolayer graphene (MLG) and $\partial_\mu n=m$ for the bilayer graphene (BLG). It can be readily checked that in both cases $\varkappa d\ll1$. 

In the absence of Galilean invariance, the mass density $\rho$ is defined as the proportionality coefficient between momentum density and hydrodynamic velocity, $\bm{p}=\rho\bm{u}$. At weak interaction we have
\[
\bm{p}=-4\int\frac{d^2p}{(2\pi)^2}\, \bm{p}(\bm{p}\cdot\bm{u})\frac{dn_F(\epsilon)}{d\epsilon}. 
\]
For the MLG this gives $\rho=9\zeta(3)T^3/(2\pi v^4)$ and $\rho=m^2T\ln 4/\pi$ for the BLG. From these estimates we can deduce that 
\begin{equation}
h=\left\{\begin{array}{cc} \frac{1}{6\pi}\sqrt{\frac{\ln8}{\zeta(3)}}\frac{vL}{Tl^2_H}, & (\text{MLG}), \\ 
\sqrt{\frac{1}{4\pi\ln4}}\frac{L}{\sqrt{mT}l^2_H}. & (\text{BLG}).\end{array}\right.
\end{equation}
Thus, we see that strong magnetoconductivity, $h\gtrsim 1$, arises at weak magnetic fields, at which the intrinsic conductivity $\sigma_0$ may be assumed independent of $H$.

To determine the temperature dependence of the intrinsic conductivity $\sigma_0(T)$ and the kinematic viscosity $\nu(T)$, we first evaluate the dimensionless interaction parameter $r_s$. 
For gate-screened Coulomb interactions in MLG, one finds $r_s\simeq e^2Td/v^2$, because the Coulomb potential $e^2/q$ saturates to $e^2d$ at momenta corresponding to distances shorter than the gate separation $d\lesssim \lambda_T$. An equivalent estimate follows from comparing the characteristic interaction energy to the kinetic energy of Dirac quasiparticles, yielding the same scaling. With this, the intrinsic conductivity is expected to scale as $\sigma_0\simeq e^2/r^2_s\propto 1/T^2$. 
The electron-electron scattering rate behaves as $\tau^{-1}_{ee}\propto Tr^2_s$, so the kinematic viscosity is of order $\nu\simeq v^2\tau_{ee}\propto1/T^3$. These considerations imply that the dimensionless parameter $\alpha\sim1$ is parametrically of order unity. 

For BLG at low temperatures, when the average spacing between the thermally excited carriers exceeds the distance to the gate, the interaction parameter is temperature-independent, $r_s\sim e^2d\Delta/v^2$. Since the ratio $T\varkappa^2/\rho$ is also temperature-independent, both the intrinsic conductivity $\sigma_0$ and the parameter $\varsigma$ are not expected to exhibit strong temperature dependence in this regime.

Using the above estimates, the temperature dependence of the fluctuation contribution to the macroscopic conductivity at $H=0$ may be evaluated as follows, 
\begin{equation}
\sigma_{\text{fl}}(T)\simeq{\sigma}_Q\left(\frac{e^2}{v}\right)^2\left\{\begin{array}{cc} (Td/v)^2\ln(L/\lambda_T), & (\text{MLG}), \\ 
(\Delta d/v)^2\ln(L/\lambda_T), & (\text{BLG}),\end{array}\right.
\end{equation}
where $\sigma_Q=e^2/(2\pi)$ is the quantum conductance. For a twisted bilayer graphene with a linear electron dispersion, we expect the same behavior as for MLG.

In summary, we generalized the classical theory of hydrodynamic fluctuations in 2D liquids~\cite{Forster:1977,Andreev:1980}, to systems with a nonzero intrinsic conductivity. 
Deviations from charge-neutrality caused by the Johnson–Nyquist noise couple the charge current to the fluctuations of the  hydrodynamic flow. This coupling produces a contribution to the macroscopic conductivity given by Eq.~\eqref{eq:sigma-H}. This contribution grows logarithmically with the system size and is very sensitive to the magnetic field. The characteristic field $H_T$ for the magnetoconductivity (MC) is given by Eq.~\eqref{eq:H_T} and corresponds to  $l_H \sim \sqrt{\lambda_T L}$. The shape of MC contains a Lorentzian feature of width $H_T$ and logarithmic dependence at $H> H_T$.

An interesting aspect of the main result, Eq.~\eqref{eq:sigma-H}, is that in the ideal liquid approximation, where all the dissipative coefficients of the electron liquid are set to zero, the fluctuation contribution to the conductivity diverges as $\sigma_{\text{fl}} \propto 1 / \sigma_0$. 
This may seem counterintuitive, since the strength of the Langevin sources vanishes at $\sigma_0 \to0$. The reason for this divergence may be understood as follows. The variance of equal-time fluctuations of the charge density, being a thermodynamic quantity, is independent of $\sigma_0$. On the other hand, the relaxation time of the  charge density fluctuations diverges at $\sigma_0 \to0$, resulting in divergent advection current.  
A similar counterintuitive dependence on the dissipative coefficients of the electron liquid occurs in the resistivity in the presence of long-range disorder~\cite{Andreev:2011} and in fluctuation-driven thermal drag  \cite{Levchenko:2022}.

The effect of weak momentum relaxation induced by disorder and electron-phonon scattering
can be incorporated into Eq.~\eqref{eq:dpdt} by introducing a friction force proportional to the fluid velocity, $-k\bm{u}_{\bm{q},\omega}$~\footnote{\textcolor{black}{For example, in the model of charge puddle disorder \cite{Lucas:2016,Li:2020}, it can be shown that $k=\frac{e^2}{2\sigma_0}\langle\delta n^2\rangle$, where $\delta n$ denotes density variations induced by the disorder potential and the angular brackets indicate an average over the system.}}. To leading order, momentum relaxation be accounted for by making the following replacement $\gamma_H\to\gamma_H+k/\rho$ in our expressions.

Finally, let us compare our mechanism of magnetoresistance (MR)  to the previously discussed mechanisms of MR  in graphene devices at charge neutrality. If a hydrodynamic flow perpendicular to the direction of the applied electric field  is allowed by the sample geometry, as is the case for Corbino samples, the EMF in the frame of the liquid is reduced by the flow, leading to positive MR~\cite{Muller:2008b,Li:2022,Gall:2023,Levchenko:2024}. In our Hall bar geometry, this mechanism is precluded by the boundary conditions. In the absence of a macroscopic flow, bulk~\cite{Narozhny:2015} and boundary~\cite{Alekseev:2015} scattering contributions to the MR at charge neutrality were discussed. 
The characteristic field strengths associated with these mechanisms are independent of the sample size. Our mechanism, produces strong MR at much weaker fields $H_T$, which are inversely proportional to the sample size, see in Eq.~\eqref{eq:H_T}. It also has a markedly different dependence on $H$. This testing  enables identifying the contributions of the different MR mechanisms and testing theoretical predictions in experimental.

We thank Matthew Yankowitz for discussions that in-part inspired this work. The work of A. L. was supported by NSF Grant No. DMR-2452658 and H. I. Romnes Faculty Fellowship provided by the University of Wisconsin-Madison Office of the Vice Chancellor for Research and Graduate Education with funding from the Wisconsin Alumni Research Foundation. The work of A. V. A. was supported by the National Science Foundation (NSF) Grant No. DMR-2424364.

\bibliography{biblio}

\end{document}